\documentclass{article}
\usepackage{spconf}

		\usepackage[utf8]{inputenc}
		\usepackage{cite}
		\usepackage{graphicx}
		\usepackage{tikz}
			\usetikzlibrary{shapes.geometric}
			\usetikzlibrary{calc}
			\usetikzlibrary{external} 
		\usepackage{tikz-3dplot}
		\usepackage{datatool}
		\usepackage{pgfplots,pgffor}
			\pgfplotsset{compat=newest,every axis/.append style={font=\small},}
		\usepackage{amsmath,amssymb,amsthm}
		\usepackage{url}
		\usepackage{etoolbox}
		\usepackage{xpatch}
		\usepackage{algorithmic}
		\usepackage{array}
		\usepackage[caption=false,font=normalsize,labelfont=sf,textfont=sf]{subfig}
	
		\usepackage{subfiles}
		\usepackage{hyperref}

	\def\reals{\mathbb{R}}
	
	\newcommand{\listint}[1]{\lbrace 1,2,\dots,#1\rbrace}
	\newcommand{\matrices}[1]{\operatorname{\mathbb{T}}\left( #1 \right)}
	\newcommand{\matricesP}[1]{\operatorname{\mathbb{T}_+}\left( #1 \right)}

	\newcommand{\changetoAlg}{ 
		\renewcommand{\figurename}{Alg.}
	}
	\newcommand{\changetoFig}{
		\renewcommand{\figurename}{Fig.}
	}
	
	\makeatletter%
	\@ifclassloaded{beamer}%
		{}%
		{
		
		}
		\makeatother%

		\def\figs{figs}
		
		\def\bib{.}
	
\title{CONVOLUTIONAL GROUP-SPARSE CODING AND SOURCE LOCALIZATION}
\name{Pol del Aguila Pla and Joakim Jaldén \thanks{Thanks to the KTH Opportunities Fund and the Royal Swedish Academy 
	  of Sciences (KVA, Stiftelsen Hierta Retzius fond and stipendierfond, call ES2017-0011) for 
	  travel funding, and to Mabtech AB for funding and data. Thanks to the Swedish Research Council
	  (VR) for funding (grant 2015-04026).} }
\address{School of Electrical Engineering and Computer Science\\
		 Department of Information Science and Engineering \\
		 KTH Royal Institute of Technology, Stockholm, Sweden\\
		  \href{mailto:poldap@kth.se}{\nolinkurl{[poldap,jalden]@kth.se}}}
		 
		\def\figs{./figs}
		\def\bib{.}		
		
\makeatletter
\g@addto@macro\normalsize{%
  \setlength\abovedisplayskip{4pt}
  \setlength\belowdisplayskip{4pt}
  \setlength\abovedisplayshortskip{4pt}
  \setlength\belowdisplayshortskip{4pt}
}
\makeatother

\begin{document}
%
\maketitle

\begin{abstract}
	In this paper, we present a new interpretation of non-negatively constrained convolutional coding problems as
blind deconvolution problems with spatially variant point spread function. 
In this light, we propose an optimization framework that generalizes
our previous work on non-negative group sparsity for convolutional models. 
We then link these concepts to source localization problems that arise in scientific imaging, 
and provide a visual example on an image derived from data captured by the Hubble telescope.
\end{abstract}

\section{Introduction}
\label{sec:intro}
		
	Convolutional sparse representations, or convolutional sparse coding (CSC), 
	have been the subject of study of many recent publications in the deep learning
	and signal processing communities, see, for example, \cite{Wohlberg2016} and the references 
	therein. CSC has been used for feature extraction in biological imaging \cite{Pachitariu2013},
	musical representation and transcription \cite{Blumensath2006,Cogliati2015}, 
	and pedestrian detection \cite{Sermanet2013}, among others. The fundamental advantage of CSC over 
	previous sparse representations is that it naturally allows for invariance constraints
	to be part of the feature extraction process, e.g. \cite{Barthelemy2012,Wohlberg2016}. 
	Besides imposing desired properties on the extracted features, these invariance constraints 
	greatly reduce the number of model parameters, and thereby, training complexity.
	
	In this paper, we shift the paradigm from feature extraction towards generative 
	models and inverse problems. In fact, we interpret convolutional coding as a means
	of relaxing the invariance constraints of a simple convolutional model. 
	A known approach to blind deconvolution with a spatially variant (SV) point spread function 
	(PSF) is approximating the SV kernel as a spatially weighted combination of elements of a low-dimensional 
	kernel basis \cite{Escande2015}. In this paper, we show that a non-negatively constrained version of the 
	optimization problem used for convolutional coding can be interpreted in this manner, placing CSC in the context
	of a longstanding open problem with many recent contributions
	\cite{Daube-Witherspoon1986,Flicker2005,Escande2015,Yun2017,OConnor2017}.
	With this new framework in mind, we propose to replace the sparsity regularizer in the convolutional
	basis pursuit denoising problem \cite{Wohlberg2016} by a non-negative group-sparsity regularizer \cite{Yuan2006}, 
	and call the resulting technique convolutional group-sparse coding (CGSC). This novel technique 
	generalizes our proposal in \cite{AguilaPla2017,AguilaPla2017a}, and can be applied to a number of 
	settings by using different groupings of the variables involved. 
	
\begin{figure}
		\centering
		\includegraphics[keepaspectratio=true,width=.995\linewidth]{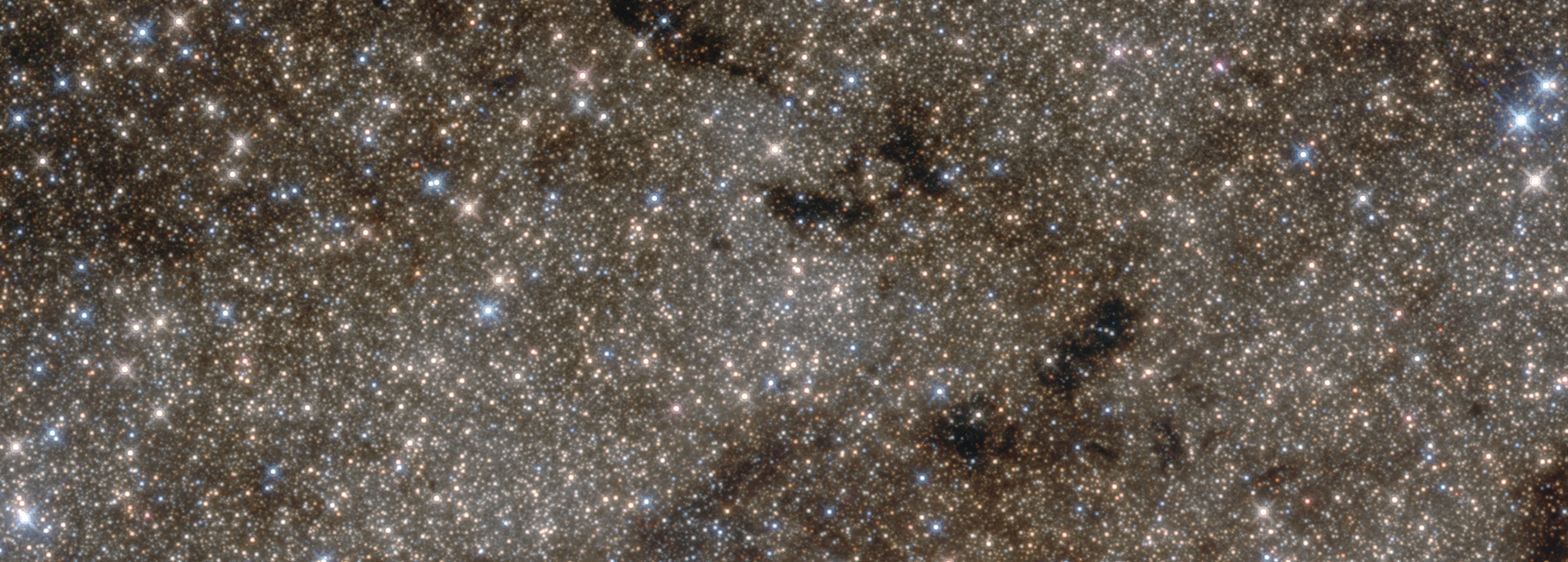}\\
		\small (a) Section of an image of the Milky Way's nuclear star cluster \\ \vspace{2pt}
		\includegraphics[keepaspectratio=true,width=.995\linewidth]{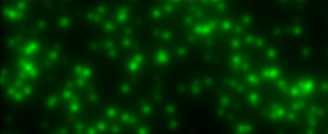} \\
		\small (b) Section of a synthetic Fluorospot observation 
		\vspace{-5pt}
		\caption{\small Scientific imaging examples in which source localization (SL) is needed. Above, section of a composite color image 
				 of the Milky Way's nuclear star cluster, generated by The Hubble Heritage Team, NASA and ESA (STScI-2016-11)
				  from an image capture using Hubble's Wide Field Camera 3. 
				 Below, a section of a synthetic Fluorospot image observation, generated using our results in \cite{AguilaPla2017a}.
				 \label{fig:Examples-SL} \vspace{-10pt}}
	\end{figure}

	For example, in diverse scientific scenarios, image data can be explained in terms of a number of 
	point- or extended-sources emitting some measurable signal (see Fig.~\ref{fig:Examples-SL} for two example images), e.g. 
	\cite{Giovannelli2005,Rebhahn2008,Smal2010,Ayasso2013,Benfenati2015,Mourya2015,AguilaPla2017,AguilaPla2017a}, and
	source localization (SL) methods automate the accurate localization of these sources.
	In \cite{AguilaPla2017,AguilaPla2017a}, we derived and analyzed a physically-motivated generative 
	model for ELISPOT and Fluorospot biomedical images. 
	This led us to an optimization problem formulation for SL on these images that naturally included a fixed convolutional
	dictionary and non-negativity constraints on the feature maps. In this paper, we generalize that formulation to fit 
	generic SL problems, and present a visual example with a composite image obtained from data captured by one of the wide
	field cameras in the Hubble telescope.

\section{Convolutional Group-Sparse Coding}
\label{sec:cgsc}
	
\subsection{Convolutional coding} \label{sec:CGSC:CC}
	In convolutional coding, one models an image observation $s\in\matrices{M,N}$ as
	\begin{equation} \label{eq:model}
		 s = \sum_{k=1}^{K} h_k \circledast x_k + n\,,
	\end{equation}
	where $n\in\matrices{M,N}$ is additive noise, $\lbrace x_k \rbrace_1^K \subset\matricesP{M,N}$
	are $K$ different images or feature maps, and $\lbrace h_k\rbrace_{1}^{K} $ is some dictionary 
	of convolutional kernels. Here, $\matrices{\mathcal{S},Q_1,Q_2}$ are the matrices of dimension 
	$Q_1\times Q_2$ with elements in the one-dimensional set $\mathcal{S}$, $\matrices{\reals,Q_1,Q_2}$
	is shortened as $\matrices{Q_1,Q_2}$, and $\matrices{\reals_+,Q_1,Q_2}$ as $\matricesP{Q_1,Q_2}$. 
	Furthermore, $\circledast$ represents a zero-padded, same-size discrete convolution.
	For the model in \eqref{eq:model}, one aims to find the most adequate 
	$x_k$s (according to some criteria) to explain $s$, for some specific collection of $h_k$s. 
	The non-negativity constraint $x_k \in \matricesP{M,N}$ is not standard in the literature, but does not 
	restrict use either. Indeed, one can incorporate pairs $\lbrace h,-h \rbrace$ to $\lbrace h_k \rbrace_1^K$
	to obtain the usual model. Our focus here, however, will be in preserving these non-negativity constraints 
	on the feature maps.
	
	In CSC, the model-fit is generally performed using the convolutional basis pursuit denoising problem 
	\cite{Wohlberg2016}, which, after the inclusion of the aforementioned non-negativity constraints, is of the form
	\begin{equation} \label{eq:c-bpdn}
		\min_{\lbrace x_k \in \matricesP{M,N} \rbrace_1^K } \left\lbrace
			\left\| \sum_{k=1}^{K} h_k \circledast x_k -s \right\|_2^2
			+ \lambda \sum_{k=1}^{K} \left\| x_k \right\|_1
		\right\rbrace\,.
	\end{equation}
	Non-negativity gains importance when we define the new variables $y\in\matricesP{M,N}$ and 
	$\alpha_k\in\matrices{[0,1],M,N}$ for $k\in\listint{K}$ such that 
	\begin{align*}
		y=\sum_{k=1}^{K} x_k, \mbox{ and } \alpha_k[i,j] =  \frac{x_k[i,j]}{y[i,j]} \mbox{ if } y[i,j] > 0\,,
	\end{align*}
	that allow us to rewrite the model in \eqref{eq:model} as
	\begin{equation} \label{eq:newinsight}
		s[\tilde{i},\tilde{j}] = \sum_{i,j} y[i,j] \sum_{k=1}^{K} \alpha_k[i,j] h_k[\tilde{i}-i,\tilde{j}-j].
	\end{equation}
	Here, for a specific pixel position $(i,j)$, the $\alpha_k[i,j]$s are convex combination coefficients that
	express the PSF at location $(i,j)$ with respect to the basis $\lbrace h_k\rbrace_{1}^{K}$.
	Therefore, \eqref{eq:newinsight} reveals that solving \eqref{eq:c-bpdn} is equivalent to 
	performing least-squares recovery of a sparse image $y$ that has been blurred by a SV PSF restricted to the 
	convex hull of $\lbrace h_k\rbrace_{1}^{K}$. 
	
	As it is common in CSC \cite{Wohlberg2016}, we will further on consider a norm constraint on 
	the $h_k$s to avoid the scaling ambiguity between filters and coefficients. In contrast to \cite[Section V]{Wohlberg2016}, that
	imposes $\left\|h_k\right\|_2=1$, we will impose the $h_k$s to have the same $1$-norm $\left\|h_k \right\|_1$.
	The model in \eqref{eq:newinsight} reveals that this is more convenient, because it corresponds to simply assuming that the local PSF
	has constant $1$-norm throughout the image.

\subsection{Convolutional group-sparse coding}

	The model in \eqref{eq:newinsight} suggests that the relation between the $x_k[i,j]$s in \eqref{eq:c-bpdn} 
	could be further exploited. For example, one may accept that in some area the local PSF is 
	described as a convex combination of $h_1$ and $h_2$, but it may well be the case that in some other area the combination of 
	these two kernels is unlikely. In other words, previous knowledge could be incorporated by locally forcing a 
	model selection between unlikely combinations and joint or group behavior between likely combinations. 
	To this end, we propose to define $G$ sets 
	$\left\lbrace\mathcal{G}_g\right\rbrace_1^G$ of indices $(i,j,k)$ such that 
	$\mathcal{G}_{g_1}\cap \mathcal{G}_{g_2}=\emptyset$ for $g_1\neq g_2$, and solve the following least-squares, 
	non-negative group-sparsity regularized problem instead,
	\begin{equation} \label{eq:CGSC}
		\min_{\lbrace x_k \rbrace_1^K} \! \left\lbrace
			\left\| \sum_{k=1}^{K} h_k \circledast x_k -s \right\|_w^2 \!\!\!\!
			+\! \lambda \sum_{g=1}^{G} \sqrt{ \sum_{(i,j,k)\in \mathcal{G}_g}\!\!\!\!\! x_k^2[i,j] }
		\right\rbrace\,,
	\end{equation}
	with $x_k\in\matricesP{M,N}$ for every $k$. Here, we use the norm $\|\cdot\|_w = \|w \odot \cdot\|_2$ (where $\odot$
	stands for the Hadamard product) to allow for a non-negative weighting $w\in\matricesP{M,N}$ that judges 
	differently prediction errors at different locations. 
	This new problem formulation \eqref{eq:CGSC}, which we name convolutional group-sparse coding (CGSC), will induce a 
	group behavior \cite{Yuan2006} on the elements $x_k[i,j]$ such that $(i,j,k)$ belong to the same $\mathcal{G}_g$.
	Note that \eqref{eq:CGSC} is a convex problem, and that it includes \eqref{eq:c-bpdn} as a specific case, in which one 
	chooses $M\times N\times K$ groups of a single index vector.
	
	Group sparsity regularization for convolutional coding has been used previously in the context of 
	multimodal imaging \cite{Wohlberg2016b,Degraux2017}, in which grouping promoted the fusion of
	information from different imaging sensors or modalities. However, these approaches do not consider non-negativity
	of the feature maps, which plays an fundamental role in SL and in the interpretation of CSC as a deconvolution 
	with SV kernels.

\subsection{Accelerated proximal gradient algorithm}	

	\begin{figure} \small
	\begin{algorithmic}[1]
	\vspace{.5pt}\hrule height 1pt \vspace{.5pt}
	\REQUIRE { \small $\lbrace x_k^{(0)} \rbrace_{1}^{K}\subset \matricesP{M,N}$, an image $s\in\matrices{M,N}$, a weight $w\in\matricesP{M,N}$ and a kernel dictionary $\left\lbrace h_k \right\rbrace_{1}^{K}$}
	\vspace{.5pt}\hrule height .5pt \vspace{.5pt}
		\item[] 
		\STATE $l\leftarrow 0$
		\FOR{ $ k=1 $ \TO $K$}
			\STATE $\displaystyle z_k^{(0)} \leftarrow x_k^{(0)}$, \label{line:iniacc}
		\ENDFOR
		\REPEAT
			\STATE $l\leftarrow l+1$
			\STATE $ \displaystyle u^{(l)} \leftarrow \sum_{k=1}^{K} h_k \circledast z^{(l-1)}_k - s$ \label{line:forwardanddiff} 
			\FOR{ $ k=1 $ \TO $K$} \vspace{2pt}
				\STATE $ \displaystyle x_k^{(l)} \leftarrow \left[ z_k^{(l-1)} - h^{\mathrm{m}}_k \circledast \left[w \odot u^{(l)}\right] \right]_+$ \label{line:adjointandpos}
			\ENDFOR
			\FOR{ $g=1 $ \TO $G$}
				\STATE $ \displaystyle n \leftarrow \sqrt{ \sum_{(i,j,k)\in \mathcal{G}_g} \left(x^{(l)}_k[i,j]\right)^2 }$ \label{line:shrinkth1}
				\FOR{ $(i,j,k)\in\mathcal{G}_g$} \vspace{2pt}
					\STATE $ \displaystyle x^{(l)}_k[i,j] \leftarrow \left(1-\frac{\lambda}{2} n^{-1}\right)_+ x^{(l)}_k[i,j]$ \label{line:shrinkth2}
				\ENDFOR
			\ENDFOR
			\FOR{ $ k=1 $ \TO $K$} \vspace{2pt}
				\STATE $z_k^{(l)} \leftarrow x_k^{(l)} + \alpha(l) \left( x_k^{(l)} - x_k^{(l-1)}\right)$ \label{line:acc2}
			\ENDFOR
		\UNTIL{ convergence of $\lbrace x_k^{(l)} \rbrace_1^K$ }
		\item[]
	\vspace{1pt}\hrule height 1pt \vspace{-10pt}
	\end{algorithmic}
	\caption{ \small 
			APG algorithm for CGSC, solving \eqref{eq:CGSC}. 
			The sequence $\alpha(l)$ can be that in \cite{Beck2009} or that in \cite{Chambolle2015}.
			\label{algs:AccProxGradforRegInvDif}  \vspace{-10pt}
			}
\end{figure}

	Proximal optimization algorithms have been widely used for CSC \cite{Chalasani2013,Cogliati2015,Gu2015,Heide2015,Wohlberg2016}.
	In \cite{AguilaPla2017}, we derived the accelerated proximal gradient (APG) algorithm (also known as
	FISTA) to solve a functional optimization problem closely related to \eqref{eq:CGSC}. Here, we will derive the APG algorithm to
	solve \eqref{eq:CGSC}, reported in Fig.~\ref{algs:AccProxGradforRegInvDif}, often referring to our previous results in 
	\cite{AguilaPla2017}. 
		
	To obtain the APG algorithm for \eqref{eq:CGSC}, one needs to characterize the mapping 
	$\lbrace x_k \rbrace_1^K \mapsto \sum_{k=1}^{K} h_k \circledast x_k$ in terms of A) its 
	adjoint and B) an upper bound on its operator norm (see \cite[Section IV-B]{AguilaPla2017}). 
	With respect to A), while in \cite{AguilaPla2017} we relied on the self-adjointness of convolutional
	operators with symmetric kernel, here we have not imposed any symmetry restrictions on the $h_k$s. 
	Nonetheless, following \cite[Appendix A, Proofs - Property 3 and Lemma 2]{AguilaPla2017},
	it is clear that the adjoint we seek is the mapping 
		$u \mapsto \left\lbrace h^{\mathrm{m}}_k \circledast \left[w \odot u\right] \right\rbrace_1^K$,
	where the $h^{\mathrm{m}}_k$s are the matched filters corresponding to the $h_k$s, constructed by
	inverting the order of the elements in the kernel matrix in both dimensions. This result was implicitly 
	stated before in \cite{Chalasani2013} in its derivation of the FISTA for CSC.
	With respect to B), in \cite{AguilaPla2017} we used that Gaussian functions have unit $1$-norm. As mentioned in Section~\ref{sec:CGSC:CC}, and to simplify
	the expressions in the final algorithm, we assume that $\left\| h_k \right\|_1 = K^{-1} \|w\|^{-2}_\infty$,
	which, following \cite[Appendix A, Proof - Lemma 1]{AguilaPla2017}, provides that
	the operator norm of the aforementioned mapping is bounded above by $1$. 
	
	Furthermore, one needs to obtain the proximal operator \cite{Bauschke2011} corresponding to the non-negative
	group-sparsity regularizer in \eqref{eq:CGSC}. In \cite[Appendix B - Lemma 6]{AguilaPla2017}, we provided the proximal
	operator of the non-negative $2$-norm in some space. The proofs there generalize well to each partition
	of $\matrices{M,N,K}$ established by the index groups $\lbrace \mathcal{G}_g\rbrace_1^G$. Combining those results with
	the separable sum property \cite{Parikh2014}, we obtain that if $\lbrace z_k\rbrace_1^K$ are the result of evaluating
	the proximal operator of the non-negative group-sparsity regularizer at $\lbrace x_k \rbrace_1^{K}$, then
		$z_k[i,j] = Q_+[i,j,k] \left( x_k[i,j] \right)_+$
	with 
	\begin{equation*}
		Q_+[i,j,k] = \left(1-\frac{\lambda}{2} \left[ 
			\sqrt{ \sum_{(i,j,k)\in \mathcal{G}_g} \left( \left[ x_k[i,j]\right]_+ \right)^2 }
		\right]^{-1} \right)_+,
	\end{equation*}
	if $(i,j,k)\in\mathcal{G}_g$ and $Q=1$ if $(i,j,k)$ does not belong to any of the $\mathcal{G}_g$s. 
	As expected, this result has the structure of a block soft thresholding operator \cite{Parikh2014}, 
	but with the addendum of the projection onto the non-negative half-space. 
	
	These results are reflected in the algorithm in Fig.~\ref{algs:AccProxGradforRegInvDif}. In particular, 
	Lines~\ref{line:forwardanddiff} and \ref{line:adjointandpos} implement the gradient step with adequate step-size and 
	non-negative projection, and Lines~\ref{line:shrinkth1} and \ref{line:shrinkth2} implement 
	the proximal operator of the non-negative group-sparsity regularizer.

\section{CGSC for source localization}
\label{sec:sl}
	
	In many cases, image-based SL problems are characterized by noisy images 
	with bright spots of different shapes and sizes that often occlude each other. Each of these spots 
	represents a relevant source, and thus, the objective of the problem is to accurately 
	distinguish and locate each of the spots. Example applications are the localization and counting
	of stars in astronomy and cells in biology, e.g. Fig.~\ref{fig:Examples-SL}.
	
	These problems can be addressed in terms of a blind deconvolution problem with SV PSF. With respect to the notation
	in \eqref{eq:newinsight}, the underlying image $y$ is a detailed geographical map of the sources that characterizes them in terms of their location and brightness,
	while the shapes and sizes of each of the spots are expressed in terms of a convex combination of kernels in the dictionary $\lbrace h_k \rbrace$.
	In this context, then, we propose  the APG algorithm for CGSC in Fig~\ref{algs:AccProxGradforRegInvDif}, with $G=M\times N \times P$ groups
	of the shape 
	\begin{equation} \label{eq:groups-sl}
		\mathcal{G}_{i,j,p} = \left\lbrace (m,n,k) : m=i,n=j,k\in\aleph_p \right\rbrace\,,
	\end{equation}	 
	with $p\in\listint{P}$.
	
	In \eqref{eq:groups-sl}, the $\aleph_p$s are sets 
	of $k$s that express which kernels $h_k$ should be considered for use jointly to express a spot shape. 
	If $\listint{K}\setminus \cup_{p=1}^{P} \aleph_p \neq\emptyset$, those $k$s correspond to kernels that are used to account for background 
	patterns or other artifacts that are not to be considered sources. This choice corresponds to a grouping effect for all those terms 
	representing a single location that should be considered jointly, placing the focus on determining whether or not a specific spatial
	location held a source. Note that the application of CGSC to SL is only possible due to the non-negativity constraints in \eqref{eq:CGSC}, as the otherwise
	unconstrained optimization proble would also try to predict dark shapes in terms of the corresponding negative kernels $\lbrace -h_k \rbrace_1^K$.
	The formulation for SL composed by \eqref{eq:CGSC} and \eqref{eq:groups-sl} generalizes our proposed approach in 
	\cite{AguilaPla2017,AguilaPla2017a} for cell detection in ELISPOT and Fluorospot data.
	An application example is provided in Section~\ref{sec:expres} using the image section from the Hubble telescope (see Fig~\ref{fig:Examples-SL}).  
	
	The challenge of choosing the base $\lbrace h_k \rbrace_1^K$ to successfully approximate spot shapes
	and background patterns remains. On one hand, data-based solutions to this problem, i.e. convolutional dictionary learning,
	have been extensively discussed in the literature, e.g., \cite[Section II-D]{Wohlberg2016} and references therein. 
	On the other hand, in \cite{AguilaPla2017,AguilaPla2017a}, we analyzed the physical model for ELISPOT and Fluorospot assays,
	and derived an observation model for the resulting images. Discretizing this observation model lead us 
	to an expression for the image observation of the form in \eqref{eq:model}, where exact expressions for 
	the $h_k$s were known, and the only source of error was discretization in itself. Finally, a first approximate solution
	can be obtained for new applications by choosing an initial base $\lbrace h_k\rbrace_1^K$ based on 
	heuristic reasoning and refining it by trial-and-error.

\section{Experimental results}
\label{sec:expres}
	
\begin{figure}
	\centering
		\includegraphics[keepaspectratio=true,width=.995\linewidth]{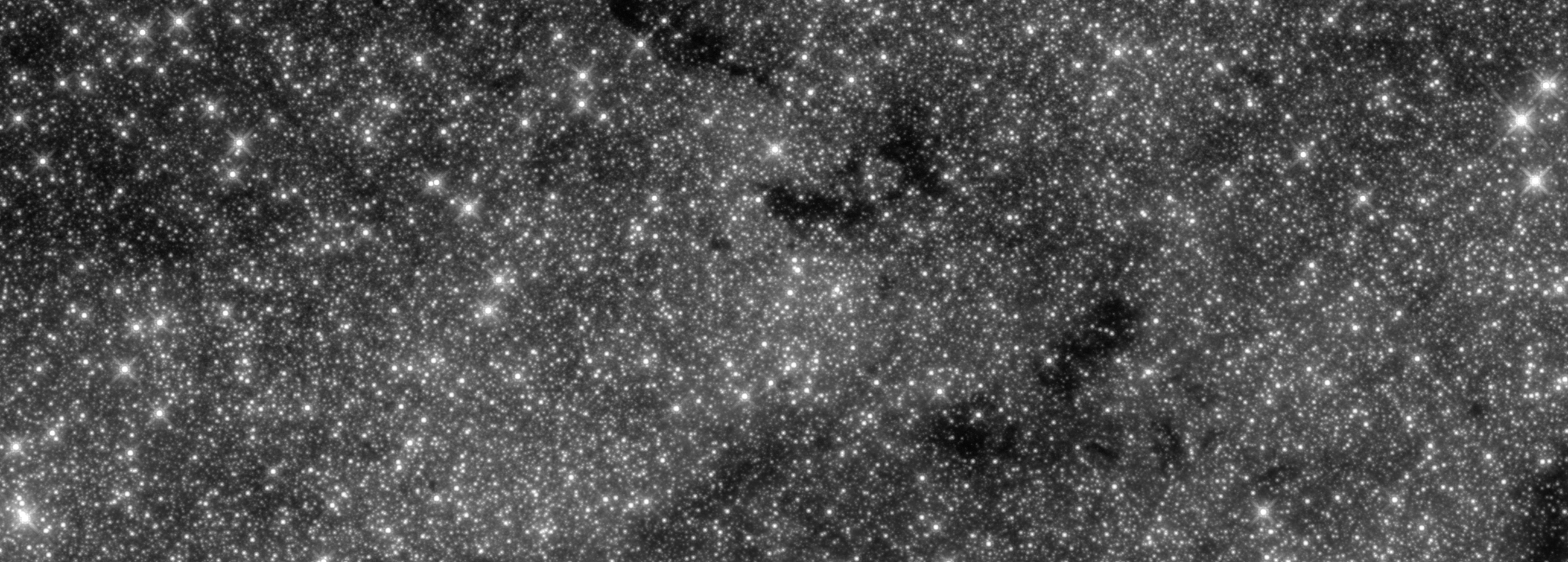}\\
		\small (a) Section of an image of the Milky Way's nuclear star cluster \\ \vspace{2pt}
		\includegraphics[keepaspectratio=true,width=.995\linewidth]{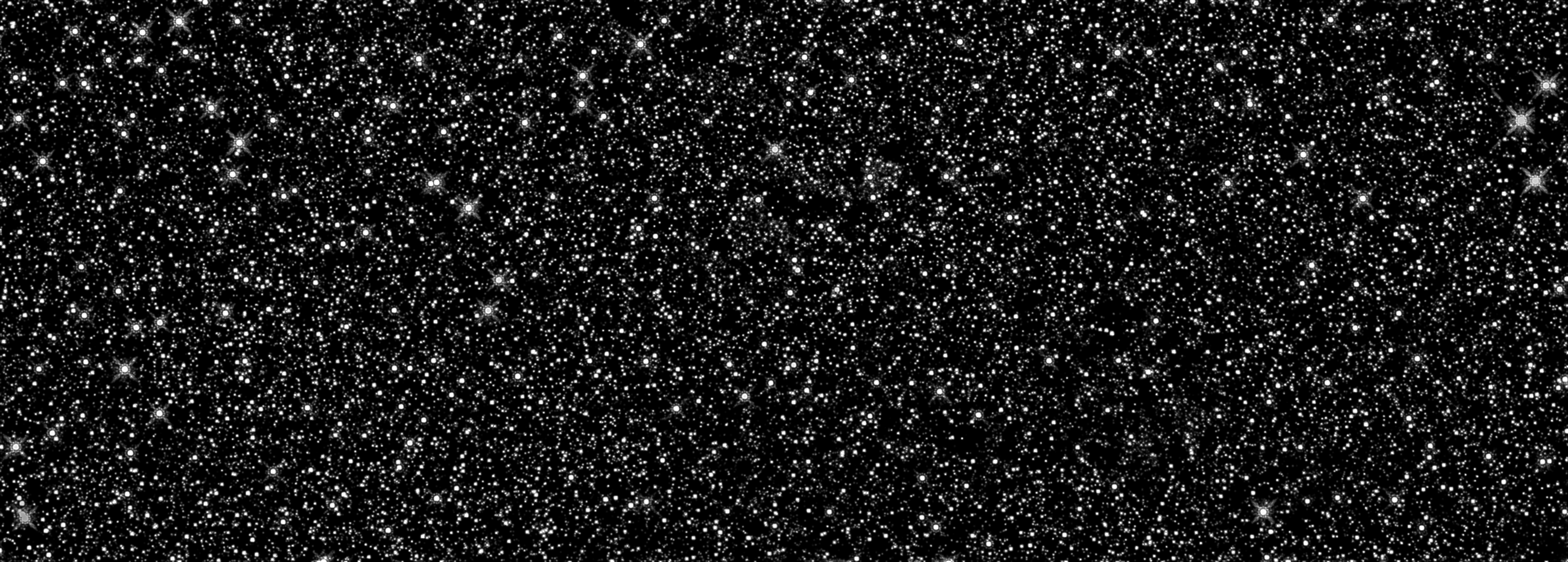} \\
		\small (b) Recovery of the foreground component using CGSC \\ \vspace{2pt}
		\includegraphics[keepaspectratio=true,width=.155\linewidth]{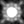}
		\includegraphics[keepaspectratio=true,width=.155\linewidth]{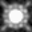}
		\includegraphics[keepaspectratio=true,width=.155\linewidth]{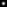}
		\includegraphics[keepaspectratio=true,width=.155\linewidth]{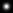}
		\includegraphics[keepaspectratio=true,width=.155\linewidth]{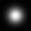}
		\includegraphics[keepaspectratio=true,width=.155\linewidth]{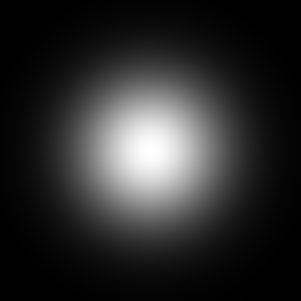}\\
		\small (c) Battery of filters $\lbrace h_k \rbrace_1^K$, $k$ increasing from left to right. 
		\vspace{-5pt}
		\caption{\small 
				 \label{fig:MilkyWay} 
				 Foreground reconstruction using the APG algorithm for CGSC in Fig.~\ref{algs:AccProxGradforRegInvDif}.
				 Above, grayscale image. Middle, foreground recovery, artificially saturated for printing clarity, showing only lower half of dynamic range. 
				 Below, set of filters used for foreground (left to right, first $5$ filters) and background prediction (last filter).
				 \vspace{-10pt}
				 }
\end{figure}
	
	In our previous works \cite{AguilaPla2017,AguilaPla2017a}, we provided extensive quantitative validation of a restricted version
	of the algorithm in Fig.~\ref{algs:AccProxGradforRegInvDif} for cell detection on ELISPOT and Fluorospot data, both with 
	synthetic images and real, expertly-labeled images.
	Here, we present a qualitative example from an entirely different setting, using the more general formulation in
	\eqref{eq:CGSC}. The image in the upper part of Fig.~\ref{fig:Examples-SL} is a section of a composite image of the Milky Way's nuclear star
	cluster, generated as part of a study to determine the structure and origin of this cluster (STScI-2016-11). The 
	study aimed to recover the cluster's total mass and its dynamic behavior. For both these purposes, an accurate localization
	and count of the stars from this and similar images was important. Besides the massive amount of stars, part of the 
	problem in this task was the background variations caused by dense dust clouds that surrounded the relevant stars and absorbed 
	part their infrared emissions. 
	
	In Fig.~\ref{fig:MilkyWay}, we report the results of a first approximation to the problem using $1000$ iterations of the 
	APG algorithm for CGSC in Fig.~\ref{algs:AccProxGradforRegInvDif}, with $\lambda=0.01$, $w=1$ everywhere and $K=6$. In particular, we choose an heuristic set 
	of $h_k$s, shown in Fig.~\ref{fig:MilkyWay}(c), and use them to predict how the image, Fig.~\ref{fig:MilkyWay}(a), would appear without
	a background component, i.e., Fig.~\ref{fig:MilkyWay}(b). 
	Of the filters in Fig.~\ref{fig:MilkyWay}(c), the first is a middle-sized star with a complex shape extracted from the image itself,
	the second is an up-sampled (scale $4/3$) version of the same star, included to match larger stars, and the remaining four are 
	spatially-integrated (within a pixel) rotationally-symmetric Gaussian kernels with different standard deviations included to match smaller stars
	(from left to right, $\sigma=1$, $\sigma=2$, $\sigma=5$) and background variations (last, $\sigma=50$). 
	Consequently, with respect to the notation in Section~\ref{sec:sl}, we choose $\aleph_1=\lbrace 1\rbrace$, $\aleph_2=\lbrace 2 \rbrace$, 
	$\aleph_3=\lbrace  3,4,5 \rbrace$. This implies that for each pixel, we consider three different groups of variables, the two
	first corresponding to the explanation of a spot by a simple copy of one of the two first kernels, and the third corresponding 
	to the explanation of a spot by a convex combination of the shapes in the next three kernels. Finally, the variable that
	corresponds to the last kernel is not regularized. Therefore, if a pattern can be matched by the last, much wider kernel, that 
	explanation will be preferred, and the pattern will be considered as part of the background.
	
	Fig.~\ref{fig:MilkyWay}(b) exhibits an impressive foreground reconstruction, at the low cost of a basic heuristic proposal 
	for $\lbrace h_k \rbrace_1^K$ based only on the most obvious patterns in the image. Specialized insight on the physical modeling of star emissions
	and the patterns that arise in the Hubble's telescope wide field cameras could, in our opinion, open the door to complete SL and counting in 
	these images at the remarkable accuracy levels we obtained in \cite{AguilaPla2017,AguilaPla2017a} for ELISPOT and Fluorospot data.

\bibliographystyle{IEEEbib}
{\small
\bibliography{\bib/multi_deconv}}

\end{document}